\documentclass[superscriptaddress,twocolumn,showpacs,prl,preprintnumbers]{revtex4}

\usepackage[english]{babel}
\usepackage{amsmath,amssymb}
\usepackage{graphicx}
\usepackage[sort&compress]{natbib}
\usepackage{xcolor}

\definecolor{red}{rgb}{1.0, 0, 0}

\makeatletter
\renewcommand{\fnum@table}{\textbf{\tablename~\thetable}}
\renewcommand{\fnum@figure}{\textbf{\figurename~\thefigure}}
\makeatother

\newcommand{\Dmq}{\Delta m^2}
\newcommand{\eVq}{\ensuremath{\text{eV}^2}}

\begin{document}

\title{Are there sterile neutrinos at the eV scale?}
\author{Joachim Kopp}          
\affiliation{Fermilab, Theoretical Physics Department,
             PO Box 500, Batavia, IL 60510, USA}
\author{Michele Maltoni}       
\affiliation{Instituto de F\'isica Te\'orica UAM/CSIC, Calle de Nicol\'as Cabrera 13-15, E-28049 Madrid, Spain} 
\author{Thomas Schwetz}        
\affiliation{Max-Planck-Institut f\"ur Kernphysik,
             PO Box 103980, 69029 Heidelberg, Germany}
\date{23 March 2011} 
\preprint{FERMILAB-PUB-11-064-T, IFT-UAM/CSIC-11-12, EURONU-WP6-11-31}

\begin{abstract}
  New predictions for the anti-neutrino flux emitted by nuclear
  reactors suggest that reactor experiments may have measured a
  deficit in the anti-neutrino flux, which can be interpreted in terms
  of oscillations between the known active neutrinos and new sterile
  states. Motivated by this observation, we perform a re-analysis
  of global short-baseline neutrino oscillation data in a framework with one
  or two sterile neutrinos. While one sterile neutrino is still not
  sufficient to reconcile the signals suggested by reactor experiments
  and by the LSND and MiniBooNE experiments with null results from
  other searches, we find that, with the new reactor flux prediction,
  the global fit improves considerably when the existence of two sterile
  neutrinos is assumed.
\end{abstract}

\maketitle

{\it Introduction.}
By now a standard paradigm of neutrino physics has emerged. A
beautiful series of experiments has established the phenomenon of
neutrino oscillations. Results from solar, atmospheric, reactor, and
accelerator neutrino experiments can be accommodated nicely by
oscillations of the three neutrinos of the Standard Model, the
so-called ``active'' neutrinos, with mass-squared differences of order
$10^{-4}$ and $10^{-3}$~\eVq, see~\cite{Schwetz:2011qt,
  GonzalezGarcia:2010er} for recent fits and references. There are,
however, a few experimental results which cannot be explained within
this framework and seem to require additional neutrinos with masses at
the eV scale~\cite{Aguilar:2001ty, AguilarArevalo:2010wv}.  Such
neutrinos cannot participate in the weak interactions due to collider
constraints, and are therefore called ``sterile'' neutrinos. 

Recently another hint for sterile neutrinos has emerged from a
re-evaluation of the expected anti-neutrino flux emitted from nuclear
reactors~\cite{Mueller:2011nm}. The new prediction is about 3\% higher
than what was previously assumed~\cite{Schreckenbach:1985ep}. If
confirmed, this result would imply that all existing neutrino
oscillation searches at nuclear reactors have observed a deficit of
electron anti-neutrinos ($\bar{\nu}_e$), which can be interpreted in
terms of oscillations at baselines of order
10--100~m~\cite{Mention:2011rk}. At typical reactor anti-neutrino
energies of few MeV, standard oscillations of the three active
neutrinos require baselines of a least 1~km. Hence, the ``reactor
anomaly'' can only be accommodated if at least one sterile neutrino
with mass at the eV-scale or higher is introduced. This is particularly
intriguing because also the long-standing ``LSND
anomaly''~\cite{Aguilar:2001ty}, as well as the more recent MiniBooNE
anti-neutrino results~\cite{AguilarArevalo:2010wv} suggest the
existence of a sterile neutrino in that mass range.  

Previous phenomenological studies~\cite{Maltoni:2007zf,
  Karagiorgi:2009nb, Akhmedov:2010vy} have been performed in a
framework in which the standard three active neutrino scenario is
amended by adding one (``3+1'') or two (``3+2'') sterile neutrinos
with masses at the eV scale. These studies came to the conclusion that
an explanation of the aforementioned anomalies within these sterile
neutrino scenarios is in conflict with various constraints from other
neutrino oscillation searches at short baselines (SBL), including also
data from reactor experiments.  In this note we revisit 3+1 and 3+2
sterile neutrino oscillation schemes in the light of the new reactor
neutrino fluxes. We argue that one sterile neutrino is still not
sufficient to describe all data, whereas a 3+2 framework is now in
much better agreement with the data.

{\it New reactor fluxes and fit of SBL reactors.}
Let us first discuss the implications of the new reactor anti-neutrino flux
prediction for reactor data alone by analyzing a set of SBL reactor
experiments at baselines $L \lesssim 100$~m~\cite{Mention:2011rk}. We
include full spectral data from the Bugey3 experiment~\cite{Declais:1994su}
at 15, 40 and 95~m and take into account the Bugey4~\cite{Declais:1994ma},
ROVNO~\cite{Kuvshinnikov:1990ry}, Krasnoyarsk~\cite{Vidyakin:1987ue},
ILL~\cite{Kwon:1981ua}, and G\"osgen~\cite{Zacek:1986cu} experiments via the
rate measurements summarized in Table~II of \cite{Mention:2011rk}.
Furthermore we include the Chooz~\cite{Apollonio:2002gd} and Palo
Verde~\cite{Boehm:2001ik} experiments at $L \simeq 1$~km. We use the
neutrino fluxes from the isotopes $^{235}$U, $^{239}$Pu, $^{238}$U,
$^{241}$Pu obtained in \cite{Mueller:2011nm} and we include the uncertainty
on the integrated flux for each isotope given in Table~I of
\cite{Mention:2011rk}, correlated between all experiments. For further
technical details see~\cite{Schwetz:2011qt}.

\begin{table}
  \begin{ruledtabular}
    \begin{tabular}{l@{\quad}ccccc}
    & $\Delta m^2_{41}$ [eV$^2$] & $|U_{e4}|$ 
    & $\Delta m^2_{51}$ [eV$^2$] & $|U_{e5}|$ & $\chi^2$/dof \\
    \hline      
    3+1 & 1.78 & 0.151 &      &       & 50.1/67\\
    3+2 & 0.46 & 0.108 & 0.89 & 0.124 & 46.5/65\\
  \end{tabular}
  \end{ruledtabular}
  \caption{Best fit points for the 3+1 and 3+2 scenarios from
    reactor anti-neutrino data. The total number of data points is 69
    (Bugey3 spectra plus 9 SBL rate measurements; we
    have omitted data from Chooz and Palo Verde,
    which are not very sensitive to the model parameters, but would
    dilute the $\chi^2$ by introducing 15 additional data points). For
    no oscillations we have $\chi^2/{\rm dof} =
    59.0/69$.} \label{tab:react-bfp}
\end{table}

\begin{figure}
  \centering 
  \includegraphics[width=0.47\textwidth]{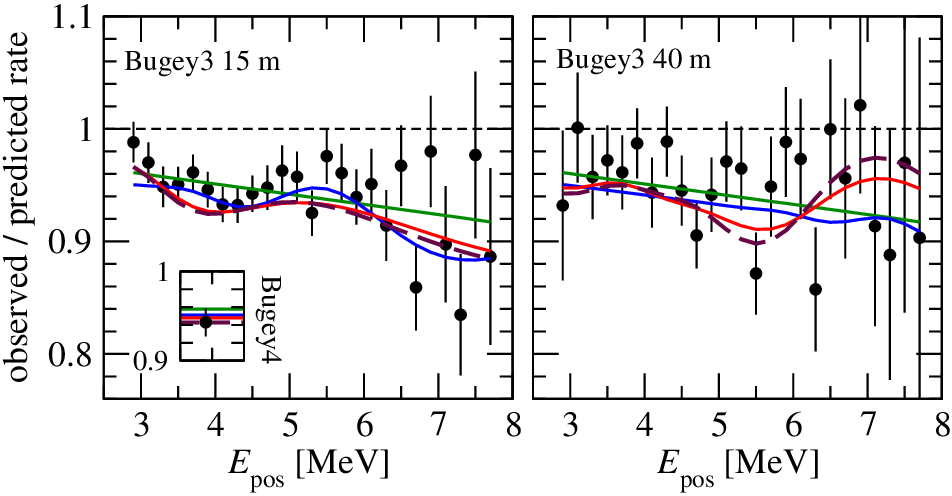}
    \caption{Comparison of sterile neutrino models to reactor data: energy
    spectra from Bugey3 and the rate measurement of Bugey4 (inset).  The
    data points correspond to the ratio of the observed event numbers to the
    predicted event number for no oscillations using the new reactor
    anti-neutrino fluxes~\cite{Mueller:2011nm}. For the Bugey3 spectra we
    show statistical errors only, whereas the error on the Bugey4 rate is
    dominated by systematics. The green solid curve shows the prediction for
    the no oscillation hypothesis, the blue solid and red solid curves
    correspond to the 3+1 and 3+2 best fit points for SBL reactor data
    (Tab.~\ref{tab:react-bfp}), and the dashed curve corresponds to the 3+2
    best fit point of global SBL data (Tab.~\ref{tab:global-bfp}).}
    \label{fig:react-data}
\end{figure}

We perform a fit to these data within the 3+1 and 3+2 sterile
neutrino frameworks, where neutrino oscillations for SBL reactor
experiments depend on 2 and 4 parameters, respectively.  The
parameters are the mass-squared differences $\Delta m^2_{41}$ and
$\Delta m^2_{51}$ between the eV-scale sterile neutrinos and the light
neutrinos, and the elements $|U_{e4}|$ and $|U_{e5}|$ of the leptonic
mixing matrix, which describe the mixing of the electron neutrino
flavor with the heavy neutrino mass states $\nu_4$ and
$\nu_5$. Obviously, for the 3+1 case, only $\nu_4$ is present.  The
best fit points for the two scenarios are summarized in
Table~\ref{tab:react-bfp}.  To illustrate the impact of sterile
neutrinos on the fit to reactor anti-neutrino data graphically, we
compare in Fig.~\ref{fig:react-data} the data to the predictions for
the no oscillation case (green) and the best fit 3+1 (blue) and
3+2 (red) models.  Note that, even for no oscillations, the
prediction may deviate from 1 due to nuisance parameters included in
the fit, parameterizing systematic uncertainties. The fit is
dominated by Bugey3 spectral data at 15~m and 40~m
and the precise rate measurement from Bugey4.

\begin{figure}
  \centering \includegraphics[width=0.47\textwidth]{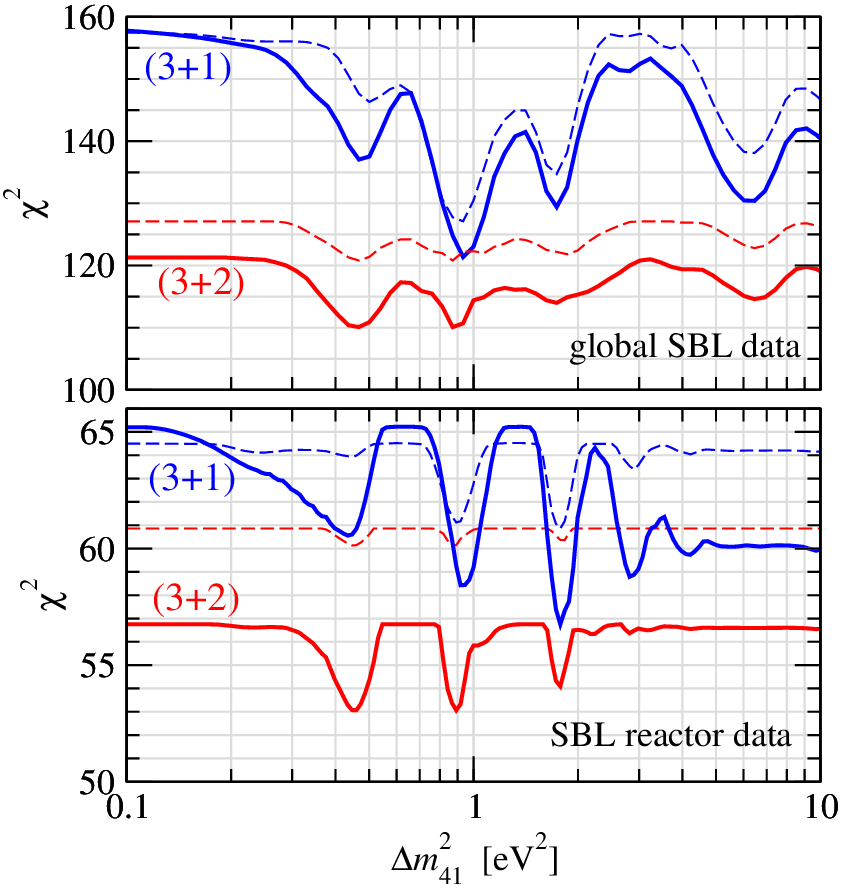}
    \caption{$\chi^2$ from global SBL data (upper panel) and from SBL
    reactor data alone (lower panel) for the 3+1 (blue) and 3+2 (red)
    scenarios. Dashed curves were computed using the old reactor
    anti-neutrino flux prediction~\cite{Schreckenbach:1985ep}, solid curves
    are for the new one~\cite{Mueller:2011nm}. All undisplayed parameters
    are minimized over.  The total number of data points is 137 (84) for the
    global (reactor) analysis.} \label{fig:chisq-dmq}
\end{figure}

In the lower part of Fig.~\ref{fig:chisq-dmq} we show the $\chi^2$ of the
SBL reactor fit as a function of $\Delta m^2_{41}$. Using the new flux
predictions (solid curves) we find a clear preference for sterile neutrino
oscillations: the $\Delta\chi^2$ between the no oscillation hypothesis and
the 3+1 best fit point is 8.5, which implies that the no oscillation case is
disfavored at about 98.6\%~CL (2~dof). In the 3+2 case the no oscillation
hypothesis is disfavored compared to the 3+2 best fit point with
$\Delta\chi^2 = 12.1$, or 98.3\%~CL (4~dof). In contrast, with previous flux
predictions (dashed curves) the improvement of the fit is not significant,
with a $\Delta\chi^2$ between the best fit points and the no oscillation
case of only 3.3 and 5.4 for the 3+1 and 3+2 hypotheses, respectively.

{\it Global analysis of SBL data.}
The constraints from the reactor experiments under discussion play an
important role in a combined analysis of all SBL oscillation data, including
the LSND and MiniBooNE anomalies. LSND has provided evidence for
$\bar\nu_\mu\to\bar\nu_e$ transitions~\cite{Aguilar:2001ty}, and MiniBooNE
has reported an excess of events in the same channel, consistent with the
LSND signal~\cite{AguilarArevalo:2010wv}. This hint for oscillations is
however not confirmed by a MiniBooNE search in the $\nu_\mu\to\nu_e$
channel~\cite{AguilarArevalo:2007it}, where the data in the energy range
sensitive to oscillations is consistent with the background expectation.
These results seem to suggest an explanation involving CP violation in order
to reconcile different results for neutrino and anti-neutrino searches. An
explanation of the LSND and MiniBooNE anomalies via sterile neutrino
oscillations requires the mixing matrix elements $|U_{e4}|$ and/or
$|U_{e5}|$ to be non-zero.  Reactor experiments are sensitive to these
parameters, and while analyses using previous flux predictions lead to tight
constraints on them, the new fluxes imply non-zero best fit values
(Tab.~\ref{tab:react-bfp}) and closed allowed regions at 98\%~CL. Hence, the
interesting question arises whether a consistent description of the global
data on SBL oscillations (including LSND/MiniBooNE) becomes now possible. To
answer this question we perform a fit by including, in addition to the
reactor searches for $\bar\nu_e$ disappearance, the
LSND~\cite{Aguilar:2001ty} and
MiniBooNE~\cite{AguilarArevalo:2010wv,AguilarArevalo:2007it} results, as
well as additional constraints from appearance
experiments~\cite{Armbruster:2002mp, Astier:2001yj}, $\nu_\mu$ disappearance
searches~\cite{Dydak:1983zq}, and atmospheric neutrinos. Technical details
of our analysis can be found in \cite{Maltoni:2007zf, Akhmedov:2010vy} and
references therein.

\begin{figure}
  \centering 
  \includegraphics[width=0.44\textwidth]{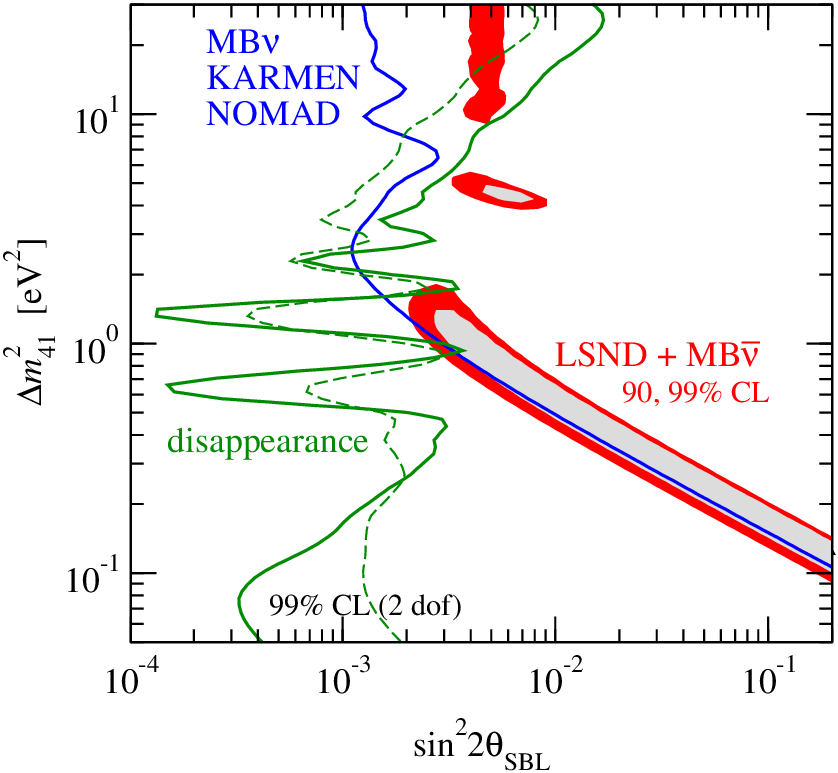}
  \caption{Global constraints on sterile neutrinos in the 3+1 model.
    We show the allowed regions at 90\% and 99\%~CL from a combined
    analysis of the LSND~\cite{Aguilar:2001ty} and MiniBooNE
    anti-neutrino~\cite{AguilarArevalo:2010wv} signals (filled
    regions), as well as the constraints from the null results of
    KARMEN~\cite{Armbruster:2002mp}, NOMAD~\cite{Astier:2001yj} and
    MiniBooNE neutrino~\cite{AguilarArevalo:2007it} appearance
    searches (blue contour).  The limit from disappearance experiments
    (green contours) includes data from CDHS~\cite{Dydak:1983zq},
    atmospheric neutrinos, and from the SBL reactor
    experiments. For the latter we compare the results for the new
    anti-neutrino flux prediction from~\cite{Mueller:2011nm} (solid)
    and the previous ones~\cite{Schreckenbach:1985ep} (dashed). The
    region to the right of the curves is excluded at 99\%~CL.}
	  \label{fig:3+1}
\end{figure}

In the 3+1 scheme the SBL experiments depend on the three parameters $\Delta
m^2_{41}$, $|U_{e4}|$, and $|U_{\mu 4}|$. Since only one mass-scale is
relevant in this case it is not possible to obtain CP violation. Therefore,
oscillations involving one sterile neutrino are not capable of reconciling
the different results for neutrino (MiniBooNE) and anti-neutrino (LSND and
MiniBooNE) appearance searches. Fig.~\ref{fig:3+1} compares the allowed regions from
LSND and MiniBooNE anti-neutrino data to the constraints from the other
experiments in the 3+1 model.  Note that, even though reactor analyses using
the new flux prediction prefer non-zero $U_{e4}$, no closed regions appear
for the disappearance bound (solid curve), since $\sin^22\theta_\mathrm{SBL}
= 4 |U_{e4}|^2 |U_{\mu 4}|^2$ can still become zero if $U_{\mu 4} = 0$. We
find that the parameter region favored by LSND and MiniBooNE anti-neutrino
data is ruled out by other experiments, except for a tiny overlap of the
three 99\%~CL contours around $\Delta m^2_{41} \approx 1$~eV$^2$. Note that
in this region the constraint from disappearance data does not change
significantly due to the new reactor flux predictions. Using the PG test
from \cite{Maltoni:2003cu} we find a compatibility of the
LSND+MiniBooNE($\bar\nu$) signal with the rest of the data only of about 
$10^{-5}$, with $\chi^2_\mathrm{PG} = 21.5 (24.2)$ for new (old) reactor
fluxes. Hence we conclude that the 3+1 scenario does not provide a
satisfactory description of the data despite the new hint coming from
reactors.

\begin{table}
  \begin{ruledtabular}
  \begin{tabular}{ccccccccc}
  & $\Delta m^2_{41}$ & $|U_{e4}|$ & $|U_{\mu 4}|$ & 
    $\Delta m^2_{51}$ & $|U_{e5}|$ & $|U_{\mu 5}|$ & 
    $\delta / \pi$ & $\chi^2$/dof\\
    \hline
    3+2 &
    0.47 & 0.128 & 0.165 & 
    0.87 & 0.138 & 0.148 & 1.64 & $110.1/130$\\
    1+3+1 &
    0.47 & 0.129 & 0.154 & 
    0.87 & 0.142 & 0.163 & 0.35 & $106.1/130$\\
  \end{tabular}
  \end{ruledtabular}
  \caption{Parameter values and $\chi^2$ at the global
    best fit points for 3+2 and 1+3+1 oscillations ($\Dmq$'s in eV$^2$).}
  \label{tab:global-bfp}
\end{table}

Let us move now to the 3+2 model, where SBL experiments depend on the
seven parameters listed in Tab.~\ref{tab:global-bfp}. In addition to
the two mass-squared differences and the moduli of the mixing matrix
elements, also a physical complex phase enters, $\delta \equiv
\mathrm{arg}(U_{\mu 4} U_{e4}^* U_{\mu 5}^* U_{e5})$. This phase leads
to CP violation in SBL oscillations~\cite{Maltoni:2007zf,
  Karagiorgi:2006jf}, allowing to reconcile differing neutrino and
anti-neutrino results from MiniBooNE/LSND.  Tab.~\ref{tab:global-bfp}
shows the parameter values at the global best fit point and the
corresponding $\chi^2$ value. Changing from the previous to the new
reactor flux calculations the $\chi^2$ decreases by 10.6 units,
indicating a significant improvement of the description of the data,
see also upper panel of Fig.~\ref{fig:chisq-dmq}. From that figure
follows also that going from 3+1 to 3+2 leads to a significant
improvement of the fit with the new reactor fluxes, which was not the
case with the old ones. The $\chi^2$ improves by 11.2 units, which
means that 3+1 is disfavoured at the 97.6\%~CL (4~dof) with respect to
3+2, compared to $\Delta\chi^2 = 6.3$ (82\%~CL) for old fluxes.

\begin{figure}
  \centering 
  \includegraphics[width=0.47\textwidth]{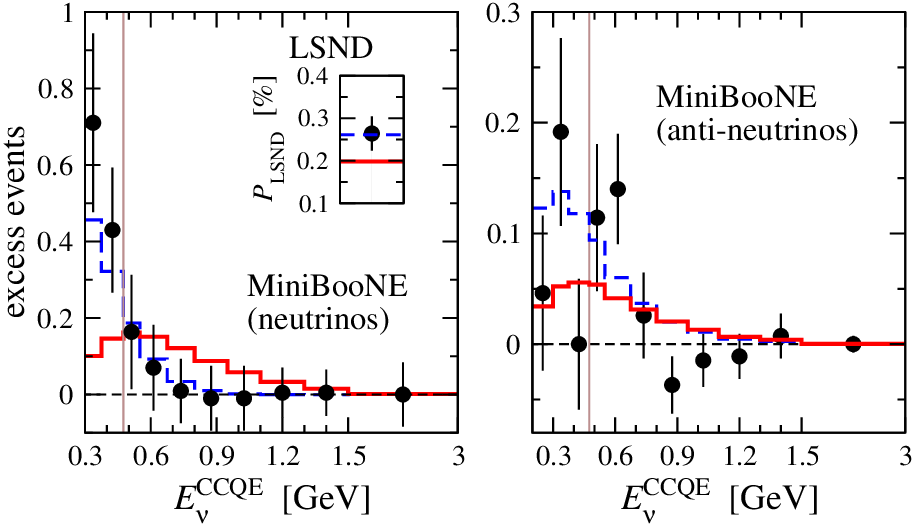}
  \caption{Predicted spectra for MiniBooNE data and the transition
    probability for LSND (inset).  Solid histograms refer to the 3+2
    global best fit point (Tab.~\ref{tab:global-bfp}), dashed
    histograms correspond to the best fit of appearance data only
    (LSND, MiniBooNE $\nu/\bar\nu$, KARMEN, NOMAD). For MiniBooNE we
    fit only data above 475~MeV.}
  \label{fig:MB-spect}
\end{figure}

\begin{figure}
  \centering 
  \includegraphics[width=0.44\textwidth]{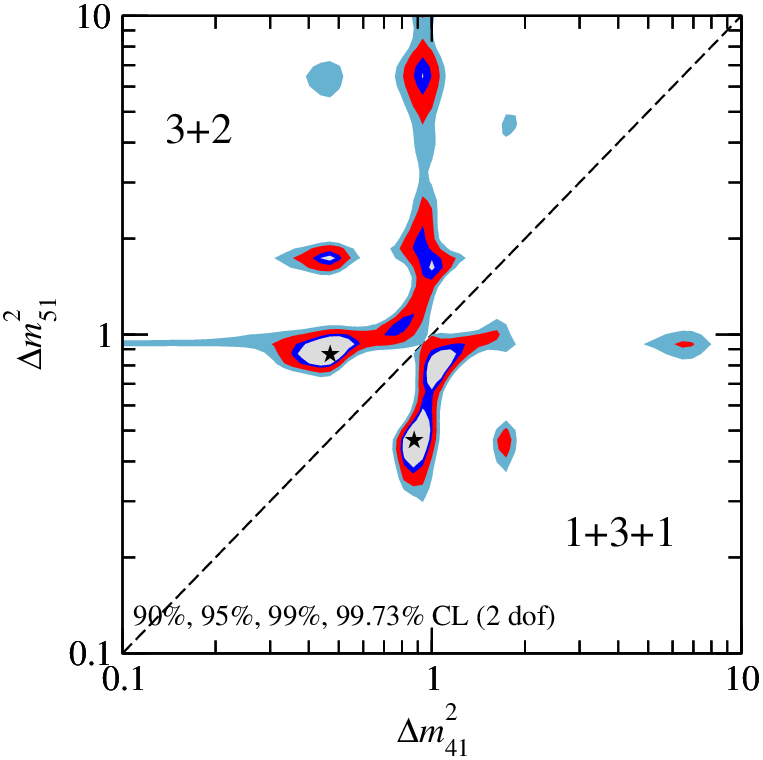}
  \caption{The globally preferred regions for the neutrino mass
    squared differences $\Delta m_{41}^2$ and $\Delta m_{51}^2$ in the
    3+2 (upper left) and 1+3+1 (lower right) scenarios.}
  \label{fig:3+2}
\end{figure}

In Fig.~\ref{fig:react-data} we show the prediction for the Bugey
spectra at the global best fit point as dashed curves. Clearly they
are very similar to the best fit of reactor data
only. Fig.~\ref{fig:MB-spect} shows the predicted spectra for
MiniBooNE neutrino and anti-neutrino data, as well as the LSND
$\bar\nu_\mu\to\bar\nu_e$ transition probability. Again we find an
acceptable fit to the data, although in this case the fit is slightly
worse than a fit to appearance data only (dashed histograms). Note
that MiniBooNE observes an event excess in the lower part of the
spectrum. This excess can be explained if only appearance data are
considered, but not in the global analysis including disappearance
searches~\cite{Maltoni:2007zf}. Therefore, we
follow~\cite{AguilarArevalo:2007it} and assume an alternative
explanation for this excess, e.g.~\cite{Hill:2010zy}.  In
Tab.~\ref{tab:PG} we show the compatibility of the
LSND/MiniBooNE($\bar\nu$) signal with the rest of the data, as well as
the compatibility of appearance and disappearance searches using the
PG test from~\cite{Maltoni:2003cu}. Although the compatibility
improves drastically when changing from old to new reactor fluxes, the
PG is still below 1\% for 3+2. This indicates that some tension
between data sets remains.  We considered also a ``1+3+1'' scenario,
in which one of the sterile mass eigenstates is lighter than the three
active ones and the other is heavier~\cite{Goswami:2007kv}. As can be
seen from Tabs.~\ref{tab:global-bfp} and \ref{tab:PG} the fit of 1+3+1
is slightly better than 3+2, with $\Delta\chi^2 = 15.2$ between 3+1
and 1+3+1 (99.6\%~CL for 4~dof). However, due to the larger total mass
in neutrinos, a 1+3+1 ordering might be in more tension with cosmology
than a 3+2 scheme~\cite{GonzalezGarcia:2010un, Hamann:2010bk,
  Giusarma:2011ex}. Fig.~\ref{fig:3+2} shows the allowed regions for
the two eV-scale mass-squared differences for the 3+2 and 1+3+1
schemes.

\begin{table}
  \begin{ruledtabular}
  \begin{tabular}{ccccc}
   & \multicolumn{2}{c}{LSND+MB($\bar\nu$) vs rest} 
   & \multicolumn{2}{c}{appearance\ vs disapp.} \\
   & old & new & old & new \\
  \hline
    $\chi^2_\mathrm{PG,3+2}$/dof& 25.1/5 & 19.9/5  
                            & 19.9/4 & 14.7/4   \\
    PG$_\mathrm{3+2}$       & $10^{-4}$ & 0.13\%  
                            & $5\times 10^{-4}$ & 0.53\% \\
  \hline
    $\chi^2_\mathrm{PG,1+3+1}$/dof& 19.6/5 & 16.0/5  
                            & 14.4/4 & 10.6/4   \\
    PG$_\mathrm{1+3+1}$     & 0.14\% & 0.7\%  
                            & 0.6\% & 3\% \\
  \end{tabular}
  \end{ruledtabular}
  \caption{Compatibility of data sets \cite{Maltoni:2003cu} for 3+2 
  and 1+3+1 oscillations using old and new reactor
  fluxes.} \label{tab:PG}
\end{table}

{\it Discussion.}
Let us comment briefly on other signatures of eV sterile neutrinos. We have
checked the fit of solar neutrino data and the KamLAND reactor experiment,
and found excellent agreement. The effect of non-zero values of $U_{e4}$ and
$U_{e5}$ for these data are similar to the one of $U_{e3}$ in the standard
three-active neutrino case, and hence the 3+2 best fit point mimics a
non-zero $U_{e3}$ close to the preferred value of these data,
see~\cite{Schwetz:2011qt, GonzalezGarcia:2010er,prep}. The MINOS
long-baseline experiment has performed a search for sterile neutrinos via
neutral current (NC) measurements~\cite{Adamson:2010wi}. We have estimated
that the best fit points reported in Tab.~\ref{tab:global-bfp} lead to an
increase of the $\chi^2$ of MINOS NC data as well as $\chi^2_\mathrm{PG}$ by
a few units~\cite{prep}. Radioactive source measurements in Gallium solar
neutrino experiments report an event deficit which could be a manifestation
of electron neutrino disappearance due to eV-scale sterile
neutrinos~\cite{Acero:2007su}. Our best fit points fall in the range of
parameter values found in~\cite{Acero:2007su} capable to explain these data.
Finally, eV-scale sterile neutrinos may manifest themselves in cosmology.
Recent studies~\cite{GonzalezGarcia:2010un, Hamann:2010bk, Giusarma:2011ex} indicate a slight
preference for extra radiation content in the universe (mainly from CMB
measurements) and one or two sterile neutrino species with masses in the
sub-eV range might be acceptable. Big-Bang nucleosynthesis leads to an upper
bound on the number of extra neutrino species of 1.2 at
95\%~CL~\cite{Mangano:2011ar}, which may be a challenge for two-sterile
neutrino schemes, or indicate a deviation from the standard cosmological
picture.

In conclusion, we have shown that a global fit to short-baseline oscillation
searches assuming two sterile neutrinos improves significantly when new
predictions for the reactor neutrino flux are taken into account, although
some tension remains in the fit. We are thus facing an intriguing
accumulation of hints for the existence of sterile neutrinos at the eV
scale, and a confirmation of these hints in the future would certainly be
considered a major discovery.

{\it Acknowledgements.} Fermilab is operated by Fermi Research Alliance
under contract~\protect{DE-AC02-07CH11359} with the US~DOE. This work is
supported by Spanish MICINN grants FPA-2009-08958, FPA-2009-09017 and
consolider-ingenio 2010 grant CSD-2008-0037, by Comunidad Autonoma de Madrid
through the HEPHACOS project S2009/ESP-1473, by the Transregio
Sonderforschungsbereich TR27 ``Neutrinos and Beyond'' der Deutschen
Forschungsgemeinschaft, and by the European Community under the EC FP7
Design Study: EURO$\nu$, Project Nr.~212372. The EC is not liable for any
use that may be made of the information contained herein.

\bibliographystyle{apsrev}
\bibliography{./3p2}

\end{document}